\documentstyle[preprint,tighten,pra,aps]{revtex}
\begin{document}
\tightenlines
\draft

\preprint{UTF 424, hep-th-9811039}

\title{Consistent, covariant and multiplicative anomalies}

\author{Guido Cognola\thanks{cognola@science.unitn.it}\ 
and  Sergio Zerbini\thanks{zerbini@science.unitn.it}}
\address{ Dipartimento di Fisica, Universit\`a di Trento \\ 
and Istituto Nazionale di Fisica Nucleare \\ 
Gruppo Collegato di Trento, Italia}

\date{January 1998}

\maketitle\begin{abstract}
It is shown that the multiplicative anomaly in the
vector-axial-vector model, which apparently 
has nothing to do with the breaking of classical current symmetries, 
nevertheless is strictly related to the well known consistent and 
covariant anomalies. 
\end{abstract}


Anomalies, as quantum breaking of classical symmetries, 
have been studied by physicists for a long time in the past, 
in the framework of a perturbative analysis based on
Feynman diagrams 
\cite{adle69-177-1426,bell69-60-47,bard69-184-1848}
and also in the context of path-integral formalism
\cite{gros72-6-477,fuji75-35-1560,fuji79-42-1195,gamb84-157-360}.

Very recently, another kind of anomaly,
the so called ``multiplicative anomaly'',
has appeared in the physical literature 
\cite{eliz98-194-613,byts98-38-1075,fili97u-23,mcke98-58-105001,dowk98u-200,evan98u-184,eliz98u-413}. 
The name, which is due mathematicians
\cite{wodz87b,kass89-177-199,kont95b},
may create some confusion between physicists, since such an anomaly
has nothing to do (at least in a direct manner) 
with the breaking of classical symmetries. 
It can be thought of as a ``measure'' of the lacking of  the commutative
property of the determinant of the product involving two 
differential operators regularised  by means of
$\zeta$-function.

Within the one-loop approximation, one meets products of operators 
in several  physical systems, namely
when one is dealing with matrix valued differential operators 
and assumes constant background fields. 
However, it is not easy to find 
situations in which the
multiplicative anomaly (from now on MA) has a direct physical relevance. 
In fact, since the MA is a local functional of the geometric data and 
of the background field, 
its contribution to the effective 
action  can be usually absorbed by the one-loop  renormalisation
procedure. Furthermore, since it comes from the 
evaluation of the one-loop functional determinant, it has a quantum origin,
i.e. it is proportional to the Planck constant and, 
as a result, it is quite easy to show that it does not give
contributions to the one loop beta-functions of the theory. 
However, this is not the case for more complicated 
systems at finite temperature and non vanishing chemical potential, where 
there exists the possibility of spontaneous symmetry breaking 
(see Ref.~\cite{eliz98-57-7430}). 
At the moment the (possible) physical importance of MA is
considered by some authors an open question (see the criticisms in Refs. 
\cite{dowk98u-200,evan98u-184,mcke98-58-105001}). 

With regard to this issue, in ref. \cite{lang96}, 
(see also \cite{lat96} in connection
with lattice anomalies), it has been shown that the non
abelian chiral anomaly can be evaluated as a regularised trace of the
commutator of pseudo-differential operators. The connection of this elegant 
and general result with the MA is that  the
latter is strictly related to the lack of ciclicity of
the regularised trace. In particular, one can see more directly why the
MA  leads to the chiral anomaly. The argument goes as
follows \footnote{We would like to thank the referee for pointing out this 
fact}. The effective action may be written as
\begin{equation}
W(A)\simeq \ln \det (D_0^+D(A))\,,
\end{equation}
where $D_0$ is a suitable free  Dirac operator and $D(A)$ is
the chiral Dirac operator mapping left spinors to right spinors.
If one makes a gauge transformation, one has
\begin{equation}
W(A_g)\simeq \ln \det (D_0^+ g^{-1}D(A)g)\,.
\end{equation}
Thus, the non vanishing of the MA leads to the loss of gauge 
invariance of 
the effective action, and this implies the existence of  the chiral anomaly. 
This kind of relevance of the MA has also been noticed in 
a completely different contest, namely within the non relativistic 
quantum statistical mechanics
\cite{mcke98-58-105001}.  

Furthermore, in the present letter we will not be able
to show that
MA is related to direct observable effects,
but nevertheless we will show again that it has an indirect physical 
importance, since it relates consistent and
covariant current anomalies in a vector-axial-vector model.
In this respect, its name is not totally unappropriated.

We recall that the current anomalies of the vector-axial-vector model 
have been derived for the
first time by Bardeen in Ref.~\cite{bard69-184-1848}.  
Later, all results have been confirmed and generalized to curved 
manifold by other authors 
\cite{reut85-31-1374,bala82-25-2713,andr84-233-232,einh84-29-331,hu84-30-836,cogn89-39-2987,cogn90-31-2699}. 

In particular, the local polynomials, 
which one has to add to the currents 
in order to pass from the consistent anomaly 
to the covariant one, have
been explicitly written down \cite{cogn90-31-2699}.

We introduce now the definition of MA. Given a positive elliptic 
 operator $D$ of positive order acting on the space of smooth sections of a
vector bundle over  
$d$-dimensional manifold without boundary, one defines
related functional  determinant by means of the associated 
$\zeta$-function. In the case of positive elliptic operators
\begin{eqnarray} 
\zeta(s|D)=\mathop{\rm Tr}\nolimits D^{-s}\:,\qquad\qquad
\det{}_\zeta D=\exp\left(-\zeta'(0|D)\right)\:.
\end{eqnarray} 
However, for first order partial operators, like the Dirac ones, one starts 
again from
\begin{equation}
\zeta(s|D)=\sum_{\lambda_n} \lambda_n^{-s}
\end{equation}
which converges if $ \mathop{\rm Re}\nolimits  s > d$, and the sum is over the whole spectrum, and 
one has to assume the existence of a spectral cut, namely 
$L_\theta=\left\{\lambda\in {\ \hbox{{\rm I}\kern-.6em\hbox{\bf C}}}, 
\theta_1<\mbox{Arg}\:\lambda<\theta_2\:\mbox{Spec}\:D\cup L_\theta=\emptyset\right\} $. Then it 
follows that there exists the analytical continuation at $s=0$ of the 
above series and again one can define \cite{kont95b}
\begin{equation}
\det{}_\zeta D=\exp\left(-\zeta'(0|D)\right)\:.
\end{equation}     
For massless chiral Dirac operators 
in even-dimensional manifold the spectral asymmetry is vanishing and one has
\begin{equation}
\ln \det D=-\frac{1}{2}\zeta'(0|D^2)\pm\frac{i\pi}{2}\zeta(0|D^2)\,,
\end{equation}
where the sign ambiguity depends on the choice of the cut of the logarithm
for the negative part of the spectrum.
This is a rigorous definition of functional determinant, well known and 
well used in quantum field theory 
\cite{dowk76-13-3224,hawk77-55-133,byts96-266-1}.
Unfortunately, it does not have all properties 
of determinant of matrices. 
If fact, for two operators, in general one has 
\begin{eqnarray} 
\zeta'(0|D_1D_2)\neq\zeta'(0|D_1)+\zeta'(0|D_2)
\end{eqnarray}
and this means that (by $\det$ we always mean $\det{}_\zeta$)
\begin{eqnarray} 
\ln\det(D_1D_2)=\ln\det D_1+\ln\det D_2+a(D_1,D_2)\:,
\end{eqnarray}
$a(D_1,D_2)$ being the MA. 
This is strictly related to the definition of determinant through
$\zeta$-function, but of course other definitions would have
other pathologies.

In simple cases, the multiplicative anomaly can be directly 
computed by definition above, but there is also a general
formula which directly follows from Wodzicki theory 
\cite{kont95b,eliz98u-413}.
If $D_1=D$ and $D_2=D+M$ are two first-order, non-commuting, 
differential operators in a (Euclidean) 
manifold without boundary, then the multiplicative anomaly 
can be written in the form \cite{eliz98u-413}
\begin{eqnarray}
a(D_1,D_2)&=& \frac{1}{4}\:\mbox{res}\:\left\{\ln^2(D_1D_2^{-1})\right\}
\nonumber\\&&\qquad
+\frac16\:\mbox{res}\:\left\{[\ln(D_1D_2^{-1})\ln D_2]^2
-\ln^2(D_1D_2^{-1})\ln^2 D_2\right\}\:, 
\label{wod0}
\end{eqnarray}
where we have omitted terms which do not contribute 
to the MA in four (or less) dimensions and $\mbox{res}(Q)$ 
is the non-commutative (or Wodzicki) residue,
which can be defined for any classical pseudo-differential operator $Q$ 
of order $q$ by the formula \cite{kass89-177-199}
\begin{eqnarray}
\mbox{res}\:(Q)=b \mathop{\rm Res}\nolimits\mathop{\rm Tr}
\nolimits (QB^{-z})|_{z=0}\:,
\end{eqnarray}
$B$ being an arbitrary elliptic operator of order
$b>q$ and $\mathop{\rm Res}\nolimits$
the usual Cauchy residue. In turns, for small z, one has
\begin{eqnarray}
{\rm Tr} (QB^{-z})\:=\frac{\mbox{res}\:(Q)}{b z}+c(Q)+O(z)\,,
\label{w2}
\end{eqnarray}
with $c(Q)$ a non-trivial term, which depends on $Q$. 

The formula for the MA above notably simplifies in the case of commuting operators. 
In particular, in two dimensions one has 
\begin{eqnarray} 
a(D,D+M)=\frac1{8\pi}\int\mathop{\rm tr}\nolimits M^2\:d^2x\:,
\label{MAn2}
\end{eqnarray} 
where $\mathop{\rm tr}\nolimits$ is the trace on internal 
or Dirac matrices.
It has to be noted that in the case of non commuting operators too,
Eq.~(\ref{MAn2}) (for dimensional reasons) 
is nevertheless the exact result, but this is true only in two
dimensions. In four dimensions (or more),
derivative terms are also present.

 Now we briefly resume the results concerning the vector-axial-vector
model.  For details and notations see 
Refs.~\cite{cogn89-39-2987,cogn90-31-2699}.    
One considers a massless Dirac particle with arbitrary internal
degrees of freedom in interaction with gauge and an axial gauge 
potentials. 

The model is described by the classical action
\begin{equation}
S=\int\left(\frac12i\overline{\psi}\gamma^k\nabla_k\psi
-\frac12i\nabla_k\overline{\psi}\gamma^k\psi
+i\overline{\psi}\gamma^k\gamma^5\!A_k\psi\right)
\:d^4x\:,
\label{azione}
\end{equation}
where
\begin{equation}
\nabla_k\psi =(\partial_k+V_k)\psi\:,
\qquad\qquad V_k=-iV_k^a\tau_a\:,
\qquad\qquad A_k=-iA_k^a\tau_a\:,
\end{equation}
are the covariant derivative, the gauge and 
the axial-gauge potentials respectively.
By $\tau_a$ we indicate  the generators of the gauge group 
and by $\gamma^k$ the (Euclidean) Dirac matrices. 

The one-loop effective
action $W$ is given
\begin{eqnarray} 
W=-\ln\det D=\zeta'(0|D)\:,
\end{eqnarray} 
with
\begin{eqnarray} 
D=i\gamma^k\nabla_k+i\gamma^k\gamma^5\!A_k\:,\qquad\qquad
D^{\dag}=i\gamma^k\nabla_k-i\gamma^k\gamma^5\!A_k\:.\qquad\qquad
\label{VAV}
\end{eqnarray}
Note that we consider the Euclidean section and so
$\gamma^k=\gamma^{k\dag}$ and $D=D^{\dag}+2i\gamma^k\gamma^5\!A_k$.

From a classical view point, both the vector and axial-vector currents
are conserved, but quantum corrections break the symmetry and 
the currents acquire anomalous terms. 
If $\delta_V$ and $\delta_A$ are the infinitesimal 
vector and axial-vector transformations,
which classically give the conserved currents, that is
\begin{eqnarray} 
\delta_V\psi=-\varepsilon\psi\:,
\qquad\qquad\delta_VV_k=\partial_k\varepsilon+[V_k,\varepsilon]\:,
\qquad\qquad\delta_VA_k=[A_k,\varepsilon]\:,
\end{eqnarray}
\begin{eqnarray}
\delta_A\psi=-\gamma^5\!\varepsilon\tau_a\psi\:,
\qquad\qquad\delta_AA_k=\partial_k\varepsilon+[V_k,\varepsilon]\:,
\qquad\qquad\delta_AV_k=[A_k,\varepsilon]\:,
\end{eqnarray}
\begin{eqnarray} 
\delta_V D=-(D\varepsilon-\varepsilon D)\:,
\qquad\qquad\delta_A D=-(D\varepsilon+\varepsilon D)\:,
\label{transD}
\end{eqnarray}
where $\varepsilon=\varepsilon^a(x)\tau_a$ is an infinitesimal matrix (depending on $x$),
then the related currents satisfy the (modified) continuity equations
\begin{equation}
\partial_k J^k +[V_k,J^k] +[A_k,\,{}^5\!J^k]={\cal A}_a\tau^a\:,
\label{vector}
\end{equation}
\begin{equation}
\partial_k\,{}^5\!J^k+[V_k,\,{}^5\!J^k] +[A_k,J^k]={}^5\!{\cal A}_a\tau^a\:.
\label{axial}
\end{equation}

In the literature, two kinds of anomalies concerning this currents
have been well studied. The first one is the so called 
consistent anomaly, which has been first derived by Bardeen from
perturbation theory and can be easily obtained by path integral,
by regularizing the one-loop effective action according to the
rule 
\begin{eqnarray} 
W_1=\frac12\lim_{s\to0}\frac{d}{ds}\:\zeta(s|D^2)\:,
\end{eqnarray}
with the implicit continuation $A_k\to iA_k$ in order to have 
a hermitian operator $D$. The gauge invariance is lost, but the Wess-Zumino 
consistency conditions are satisfied (see, for example, 
\cite{reut85-31-1374,fuji79-42-1195}). 
In this case only the axial symmetry is broken.

Using Eqs.~(\ref{transD}) one easily obtains
\begin{eqnarray}
\delta_V\mathop{\rm Tr}\nolimits D^{-2s}=0\:,
\qquad\qquad
\delta_A\mathop{\rm Tr}\nolimits D^{-2s}=-2s\mathop{\rm Tr}\nolimits(D^{-2s}\gamma^5\varepsilon)\:.
\end{eqnarray}
Since the analytic continuation of 
$\mathop{\rm Tr}\nolimits(D^{-2s}\gamma^5\varepsilon)$ 
is regular at $s=0$,
for the  consistent anomalies we have
\begin{eqnarray}
{\cal A}_a^{cons}&=&\frac{\delta_V}{\delta\varepsilon^a}\:W_1=0\:,
\label{a1}\\
{}^5\!{\cal A}_a^{cons}&=&\frac{\delta_A}{\delta\varepsilon^a}\:W_1
= \frac{1}{8\pi^2}\mathop{\rm tr}\nolimits[\tau_a
\gamma^5\! a_2({x|D}^2)]\:,
\end{eqnarray}
where $a_2(x|H)$ is the local spectral coefficient relative to the
second order operator $H$. 
It is understood that  in arbitrary even dimensions $2N$,
in the latter equation one has to replace $a_2/(4\pi)^2$ with $a_N/(4\pi)^N$. 

The second one, which is called ``covariant anomaly'', 
can be formally obtained from the ``covariant functional''
\begin{eqnarray} 
W_2=\frac12\lim_{s\to0}\frac{d}{ds}\:\zeta(s|D^{\dag}D) \:.
\end{eqnarray}
In this last case, both the currents 
$J^k$ and ${}^5\!J^k$ acquire an anomalous ``covariant'' term,
but the Wess-Zumino consistency conditions are not satisfied.
This is due to the fact that, in the $\zeta$-function regularisation,
one has to work with an arbitrary argument $s$ of $\zeta$ and only at the end
of the calculation one has to take the limit $s\to0$. 
Such a ``breaking of the consistency'' is strictly related to the operator 
one has to deal with and to the necessity of workig with regularised 
quantities. 
With regard to this, the form of the variation of the operator is crucial.

Now, again using Eqs.~(\ref{transD}) and the ``intertwining identity''
\begin{eqnarray}
D\:f(D^{\dag}D)\:D^{\dag}=DD^{\dag}\:f(DD^{\dag})\:,
\end{eqnarray}
one gets
\begin{eqnarray}
\delta_V\mathop{\rm Tr}\nolimits(D^{\dag}D)^{-s}&=&-2s
\mathop{\rm Tr}\nolimits\left\{\left[(D^{\dag}D)^{-s}
-(DD^{\dag})^{-s}\right]\varepsilon\right\}\:,\nonumber\\
\delta_A\mathop{\rm Tr}\nolimits(D^{\dag}D)^{-s}&=&-2s
\mathop{\rm Tr}\nolimits\left\{\left[(D^{\dag} D)^{-s}
+(DD^{\dag})^{-s}\right]\gamma^5\varepsilon\right\}\:,
\end{eqnarray}
The traces in the formulae above have regular analytic continuations
to $s=0$ and the first derivative with respect to $s$ at $s=0$
gives the following formulae for the covariant anomaly:
\begin{eqnarray}
{\cal A}_a ^{cov}&=&\frac{1}{16\pi^2}
\mathop{\rm tr}\nolimits\{\tau_a[a_2(x|D^{\dag}D)-a_2(x|DD^{\dag})]\}\:,
\\ 
{}^5\!{\cal A}_a^{cov}&=&\frac{1}{16\pi^2}\mathop{\rm tr}\nolimits 
\{\tau_a\gamma^5\![a_2(x|D^{\dag}D)+a_2(x|DD^{\dag})]\}\:.
\label{an2}\end{eqnarray}

Explicit expressions  for the anomalies have been computed using 
different methods also in curved space-time with 
torsion \cite{cogn89-39-2987,cogn90-31-2699}).  
For a non abelian gauge group the difference reads
\begin{eqnarray} 
{\cal A}_a^{cov}-{\cal A}_a^{cons}&=&
{\cal A}_a^{cov}=\frac{1}{8\pi^2}
\mathop{\rm tr}\nolimits[F^*\!G+G^*\!F]\:,\\ 
{}^5\!{\cal A}_a^{cov}-{}^5\!{\cal A}_a^{cons}&=&
\frac{1}{12\pi^2}\mathop{\rm tr}\nolimits\left\{G^*\!G\right.
-8e_{ijrs}
A^i A^j A^r A^s
\nonumber\\&&\left.\qquad\qquad
+4[^*\!F_{ij}A^iA^j+A^iA^j{^*}\!F_{ij}
+A^i{^*}\!F_{ij}A^j] 
\right\}\tau_a\:,
\end{eqnarray}
where 
\begin{eqnarray} 
F_{ij}&=&\partial _iV_j-\partial _jV_i+[V_i,V_j]+[A_i,A_j]\:,
\\ 
G_{ij}&=&\partial _iA_j-\partial _jA_i+[V_i,A_j]-[V_j,A_i]\:,
\end{eqnarray}
\begin{eqnarray}
^*F^{ij}\!=e^{ijrs}\:F_{rs}\:,\qquad\qquad
G^*\!G=e^{ijrs}G_{ij}G_{rs}\:.
\end{eqnarray}

It is possible to pass from the consistent anomaly to the covariant
one by adding local polynomials to the currents. Such polynomials 
$\chi^k=\chi^k_a\tau^a$ and ${}^5\!\chi^k={}^5\!\chi^k_a\tau^a$ 
are chosen to satisfy \cite{cogn90-31-2699} 
(for a misprint, in that paper 
$\chi^k$ and ${}^5\!\chi^k$ are exchanged)
\begin{eqnarray} 
&&
\partial_k\chi^k+[V_k,\chi^k]+[A_k,{}^5\!\chi^k]=
({\cal A}_a^{cov}-{\cal A}_a^{cons})\tau^a\:,
\label{diff1}\\&&
\partial_k{}^5\!\chi^k+[V_k,{}^5\!\chi^k]+[A_k,\chi^k]=
({}^5\!{\cal A}_a^{cov}-{}^5\!{\cal A}_a^{cons})\tau^a\:.
\label{diff2}
\end{eqnarray}
In this way, if $J^k$ and ${}^5\!J^k$ satisfy continuity equations
with the consistent anomalous term then 
$J^k+\chi^k$ and ${}^5\!J^k+{}^5\!\chi^k$ satisfy continuity equations
with the covariant anomalous term. The polynomials one has to add 
to the currents are chosen ``ad hoc'' in order to satisfy the previous 
conditions. Here we show that they are related to the 
corresponding multiplicative anomaly 
(connections between current terms and the Wodzicki residue have been
pointed out also in Ref.~\cite{mick93b}). 

To this aim, we first observe that
\begin{eqnarray}
\zeta'(0|D^{\dag}D)=
\zeta'(0|D^{\dag})+\zeta'(0|D)+\:a(D^{\dag},D)\,.
\label{op}
\end{eqnarray}
Moreover, since $D$ and $D^{\dag}$ only differs in the sign of $A_k$,
we also have 
\begin{eqnarray}
\delta_{V}\zeta'(0|D^{\dag})&=&-\delta_{V}\zeta'(0|D)=0\:,\\ 
\delta_{A}\zeta'(0|D^{\dag})&=&\delta_{A}\zeta'(0|D)=\frac12\delta_{A}\zeta'(0|D^2)
\label{deze}\end{eqnarray}
and finally, using Eqs.~(\ref{op}-\ref{deze}) and remembering the
definitions of the anomalies, we obtain the  result
\begin{eqnarray} 
&&
({\cal A}_a^{cov}-{\cal A}_a^{cons})
=\frac{\delta_V}{2\delta\varepsilon^a}\:a(D^{\dag},D)\:,
\\&&
({}^5\!{\cal A}_a^{cov}-{}^5\!{\cal A}_a^{cons})
=\frac{\delta_A}{2\delta\varepsilon^a}\:a(D^{\dag},D)\:.
\end{eqnarray}

The MA is a functional of $V_k$ and $A_k$, then
by performing the variations and comparing 
the result with Eqs.~(\ref{diff1})
and (\ref{diff2}) we obtain
\begin{eqnarray} 
\chi^k_a=\frac12\:\frac{\delta a(D^{\dag},D)}
{\delta V^a_k}\:,
\qquad\qquad
{}^5\!\chi^k_a=\frac12\:\frac{\delta {\cal A}(D^{\dag},D)}
{\delta A^a_k}\:.
\end{eqnarray} 

In order to  check the above relations, we have rigorously proved
using $\zeta$-function techniques, one has to evaluate MA  
by using the more general and complicated
formula, Eq.~(\ref{wod0}). This can be done in principle, but, since 
one is dealing with non commuting operators, it is not an easy task
in general, but in two dimensions, where one has the simple formula, Eq.~(\ref{MAn2}). 

Thus, let us  consider  the
simple case of abelian vector-axial vector field in two dimensions.
Such a model is described by a classical  action similar to the one in 
Eq.~(\ref{azione}), with $\gamma^0=\sigma_1$, 
$\gamma^1=\sigma_2$ and $\gamma^5=\sigma_3$.
Since we are in two dimensions now we have
\begin{eqnarray} 
{}^5\!{\cal A}_a^{cov}-{}^5\!{\cal A}_a^{cons}&=&
\frac1{4\pi}\mathop{\rm tr}\nolimits\left\{\gamma^5\left[
a_1(x|D^{\dag}D)+a_1(x|DD^{\dag})-2a_1(x|D^2)\right]\right\}
\nonumber\\ 
&=&-\frac1\pi\:\partial_k A^k\:.
\end{eqnarray}
With regard to the MA, since we are in two dimensions, we can use 
Eq.~(\ref{MAn2}) with $M=2i\gamma^k\gamma^5\!A_k$   and the result
reads
\begin{eqnarray}
a(D^{\dag},D)=\frac1\pi\int A_kA^k\:d^2x\:,
\end{eqnarray}
from which, the relation 
\begin{eqnarray} 
{}^5\!{\cal A}_a^{cov}-{}^5\!{\cal A}_a^{cons}
=\frac{\delta_A}{2\delta\varepsilon}a(D^{\dag},D)
=-\frac1\pi\:\partial_k A^k\:.
\end{eqnarray}
easily follows.

In this letter, we have shown that MA in the vector-axial-vector model
relates consistent and covariant anomalies
and we have also determined,
as functional derivatives of MA,
the local polynomials which permit
to pass from one form of the anomaly to the other one.
We have also checked the general results in a simple 
two dimensional case, where explicit computations can be performed.

\section*{Acknowledgments}{We would like to thank L.~Vanzo for useful discussions.}

\end{document}